# EGAMA-RC: Risk-Calibrated Evidence-Gated Adaptive Malware Analysis for Robust and Interpretable Memory-Forensic Triage

Isaac K. Nti, *Member, IEEE*

***Abstract*—Machine-learning malware detectors often achieve high clean-data accuracy, but operational triage also requires evidence about uncertainty, novelty, robustness, interpretability, latency, and review cost. This paper presents EGAMA-RC, a risk-calibrated evidence-gated framework for memory-forensic malware triage. Building on SHAP-guided feature refinement, EGAMA-RC combines dataset-specific refinement, model-pool evaluation, adversarial and open-family testing, novelty scoring, explanation-conditioned evidence, and runtime-aware routing. Low-risk samples are accepted automatically, while uncertain, high-risk, or potentially novel cases are routed to review, escalation, or novelty-aware handling. Across three malware datasets and a frozen multi-seed protocol, the selected hybrid gate accepts 93.12% of pooled samples with 99.86% accepted accuracy and a 0.136% false-accept rate. Novelty calibration reduces over-restrictive review behavior while preserving a low unsafe-accept profile. XGBoost provides lightweight fast-path inference with p50/p95 latency of 0.0054/0.0059 ms per sample. The results show that dependable malware analysis requires risk-calibrated routing, novelty awareness, and controlled analyst review, not classification accuracy alone.**

***Index Terms*—Memory forensics, malware detection, explainable artificial intelligence, SHAP-guided feature refinement, adversarial machine learning, selective prediction, novelty detection, runtime triage.**

## I. INTRODUCTION

Malware detection is challenging when dangerous actions are masked by obfuscation, deferred execution, packaging, code injection, or artifacts that exist only in runtime and may not be detected by static file signatures [1], [2], [3], [4], [5], [6], [7]. Memory-forensic analysis offers a complementary perspective because it can capture evidence from the volatile execution state, including processes, services, handles, loaded modules, thread activity, and injection-related traces [1], [5], [8], [9], [10], [11], [12], [13]. In SHapley Additive exPlanations (SHAP)-guided feature refinement work [1], memory-resident indicators from the CIC MalMem-2022 dataset were used to build a compact malware detection pipeline. That study examined 55 memory-resident features and reduced them to 13 forensic indicators using mutual information and SHAP-guided feature refinement (SHAP-GFR), with the retained indicators dominated by service, handle, thread, module-loading, and injection-related evidence [1].

However, strong clean-data performance alone does not establish operational reliability [5], [8], [14], [15]. A detector used for memory-forensic triage must also remain dependable under perturbation, obfuscation, family shift, latency constraints, explanation instability, and analyst-review cost [16], [17], [18], [19], [20], [21], [22], [23], [24], [25], [26], [27], [28], [29], [30]. The SHAP-GFR study [1] is an excellent illustration of this tension: The random forest achieved the highest clean Independent and Identically Distributed (IID) F1-score. XGBoost achieved similar clean performance with the lowest median per-sample latency. While Convolutional Neural Network (CNN)-based models were more stable under direct gradient-based perturbations, RF and XGBoost showed the strongest parsed-family Leave-One-Family-Out (LOFO) behavior [1]. These findings suggest that model choice depends on the deployment condition rather than on clean accuracy alone.

Current malware-detection and Explainable Artificial Intelligence (XAI) experiments [1], [2], [3], [4], [6], [10], [31], [32], [33], [34], [35] show high classification metrics and subsequently provide post-hoc explanations to understand the model behavior. That design is useful, but it leaves a gap between explanation and operational decision control. In operational triage, the system must decide not only which label to assign, but also whether the evidence is strong enough for automatic acceptance, whether the case should be escalated for higher-assurance handling, whether it resembles a novel or shifted family, or whether analyst review is required.

Prior malware-defense work typically defines an explicit operational or adversarial mechanism, such as stateful query-defense indicators in MalProtect or interpretable structural rectification in IADGuard, rather than relying only on standard classifier comparison [2], [4]. MalProtect [2], for example, uses multiple threat indicators to detect query attacks and reports evasion-rate reduction under several attacker scenarios, including adaptive settings [2].

To address this gap, we propose a risk-calibrated evidence-gated adaptive malware analysis for robust and interpretable memory-forensic triage, abbreviated EGAMA-RC. EGAMA-RC builds on the compact SHAP-GFR memory-forensic evidence layer [1] but adds a runtime decision architecture. Ref. [1] showed that no single model family dominates all operational conditions. Tree ensembles are highly accurate and efficient under clean IID and family-generalization settings, while convolutional neural models can be more resilient under gradient-based perturbations. Therefore, EGAMA needs an escalation route that moves

School of Information Technology, University of Cincinnati, Ohio, USA.
(e-mail: ntiik@ucmail.uc.edu). ORCID: 0000-0001-9257-4295

uncertain, high-risk, adversarially sensitive, or model-disagreement cases away from automatic acceptance and toward stronger review or specialist evaluation. The proposed framework uses a fast path for low-risk samples, a calibrated acceptance rule for automated decisions, escalation for higher-assurance handling when uncertainty, disagreement, attribution instability, or perturbation sensitivity is high, novelty or family-shift scoring when samples deviate from known forensic profiles, and analyst review when the evidence does not support reliable automated acceptance. In this framing, SHAP-derived evidence is not treated merely as a visualization artifact; rather, explanation stability and forensic evidence consistency become part of the routing evidence used to determine whether a prediction should be accepted, escalated, reviewed, or flagged as novel. We use the term triage to mean evidence-based decision routing rather than classification alone. A triage system must determine not only whether a sample is benign or malicious, but also whether the prediction is sufficiently reliable for automatic acceptance or should instead be reviewed, escalated, or flagged as potentially novel.

The proposed threat and risk framing of the EGAMA-RC framework follows the adversarial machine-learning taxonomy in National Institute of Standards and Technology (NIST) artificial intelligence (AI) 100-2e2023 [36]. NIST organizes adversarial ML risk around system type, lifecycle stage, attacker goals, attacker capabilities, and attacker knowledge [36]. In EGAMA-RC, the primary objective of the attacker is to cause an integrity violation, specifically by accepting a malicious memory sample as benign. Depending on the evaluation setting, the attacker may manipulate feature-space artifacts, exploit obfuscation, induce family shift, or interact with a deployed triage interface in a black-box or gray-box manner. The defender's objective is not only to maximize classification accuracy, but also to reduce false acceptance of malware and route suspicious samples to robust evaluation or analyst review. At the same time, the defender must preserve low-latency inference when the evidence supports safe automated acceptance. The contributions of this work are therefore formulated as methodological contributions:

- We propose EGAMA-RC, a risk-calibrated, evidence-gated malware-triage framework that extends memory-forensic malware detection from unconditional classification to operational decision routing. EGAMA-RC assigns each sample to automatic acceptance, analyst review, escalation, or novelty-aware handling using calibrated evidence rather than a fixed classifier output alone.
- We develop a dataset-aware forensic evidence layer grounded in SHAP-guided feature refinement. For CIC-MalMem-2022, the evidence layer builds on the validated SHAP-GFR pipeline by organizing memory-forensic indicators into interpretable groups covering service behavior, handle usage, thread/process activity, module loading, injection artifacts, and other runtime memory evidence. For BODMAS and BCCC, the framework preserves dataset-native tabular feature spaces while applying leakage-controlled refinement and audit procedures.
- We implement a selective prediction and evidence-gated routing policy based on confidence, normalized entropy, prediction margin, model-pool probability variance, inter-model disagreement, conformal-style calibration, severe-risk triggers, and family-novelty evidence. Attribution stability, perturbation sensitivity, and runtime overhead are retained as diagnostic audit evidence rather than treated as causal proof.
- We evaluate EGAMA-RC across three malware datasets; CIC-MalMem-2022, BODMAS, and BCCC, under clean IID testing, multi-seed feature refinement, model-pool comparison, evidence-gated routing, novelty-aware calibration, robustness stress testing, parsed-family/open-family analysis, attribution-stability diagnostics, and deployment-cost assessment.
- We show that hybrid evidence gating provides the strongest safety–throughput trade-off in the evaluated setting. Across the pooled evaluation, the selected hybrid gate accepts 93.12% of samples with 99.864% accepted-set accuracy, 0.136% false-accept rate, and 0.106% false-accept malware rate, while preserving review and escalation pathways for higher-risk cases.

The rest of the paper is organized as follows. Section II provides background on memory-forensic malware analysis, XAI for malware detection, adversarial robustness, concept drift, open-set malware detection, calibration, and selective triage. Section III describes the EGAMA-RC methodology, including the evidence layer, model pool, risk-calibrated gate, routing policy, explanation-conditioned evidence, threat model, latency protocol, and reproducibility plan. Section IV contains the results of our experiment and discussion, and the study concludes with future directions in Section V.

## II. RELATED WORKS

### *A. Memory-Forensic and Dynamic Malware Analysis*

Malware analysis has traditionally relied on static, dynamic, or hybrid evidence [1], [3], [37]. Static analysis is efficient but may fail when malware uses packing, obfuscation, polymorphism, or delayed behavior [18], [35], [38]. Dynamic and memory-based analysis can reveal runtime evidence, including active processes, loaded modules, registry activity, network behavior, and memory-resident injection traces[1], [18], [35], [38]. Recent works [2], [6], [7], [18] have moved toward richer runtime or multi-dimensional evidence. For example, [18] combines API, registry, network, and memory analysis to detect seen and never-seen-before malware. They reported strong performance on known binaries and more than 91% accuracy for never-seen-before binaries [18]. This confirms that memory and runtime artifacts are valuable for robust malware analysis.

Other works [11], [32], [37], [38] have explored static multi-feature representations for Internet of Things (IoT) malware detection. In [37], they extracted bytes, opcodes, API calls, strings, and DLLs, then converted them into five feature-image views, and used a multi-SPP network to detect malware variants [37]. Similarly, Ref. [3] addresses malware detection in IoT networks through trust-based user-edge evaluation and privacy-preserving feature generation [3]. These studies improve feature richness, deployment setting, or detection performance. However, they do not directly address the triage question at the center of EGAMA-RC: when should a memory-forensic prediction be accepted automatically, escalated, or withheld for analyst review? Unfortunately, these are key questions centered on alleviating false alarms in the malware detection system, which have not yet been adequately addressed in past studies.

The SHAP-GFR study [1] is positioned within this memory-forensic line of work. It uses CIC MalMem-2022 [39], reduces 55 memory-resident features to 13 SHAP-GFR indicators, and evaluates five model families under clean IID, efficiency, adversarial perturbation, and parsed-family LOFO protocols [1]. EGAMA-RC extends this foundation from model evaluation to runtime evidence-gated triage.

### *B. Explainable AI for Malware Detection*

Explainable AI has become important in malware analysis because analysts need evidence that connects model outputs to recognizable forensic behavior [1]. Surveys of XAI for malware analysis emphasize the need for explanations that support analyst reasoning, transparency, and model validation, while also noting that malware explanations can be fragile and may expose new attack surfaces [5], [6], [8], [40]. SHAP and Local Interpretable Model-Agnostic Explanations (LIME) are frequently used to identify influential features in malware classifiers, but post-hoc explanations should not be interpreted as causal proof.

In [1], SHAP and LIME were used to identify service, handle, module-loading, and injection-related indicators as important memory-forensic evidence [1]. However, SHAP and LIME explanations were primarily interpretive and feature-refinement oriented. In EGAMA-RC, we propose a different role for explanation: explanation profiles and explanation stability become runtime evidence signals. A sample whose prediction depends on unstable or inconsistent memory-artifact explanations are routed differently from a sample whose prediction is both confident and explanation-stable.

This distinction is central. EGAMA-RC does not claim that SHAP proves malicious causality. Instead, it treats explanation stability, top-feature consistency, and forensic-group agreement as reliability signals that must be validated through perturbation, deletion, or insertion fidelity tests, and agreement with known memory-forensic semantics. The methodological claim is therefore not “XAI makes the model trustworthy,” but rather “explanation-conditioned evidence can support risk-aware routing if its stability and fidelity are measured.”

### *C. Robust Malware Detection and Adversarial Machine Learning*

Machine Learning (ML)-based malware detectors are vulnerable to adversarial manipulation. NIST AI 100-2e2023 frames adversarial ML around attacker objectives, capabilities, knowledge, lifecycle stage, and attack type [36]. In malware detection, adversarial evaluation must be handled carefully because, unlike images, malware artifacts are constrained by functionality preservation and domain-specific feature semantics. MalProtect highlights this point directly, noting that malware adversarial examples differ from other domains because malware feature vectors are constrained and malicious functionality must be preserved [2].

Recent papers [2], [4] show the level of adversarial rigor expected in this area. MalProtect proposes a stateful defense against adversarial query attacks and uses several threat indicators rather than relying only on similarity or out-of-distribution detection [2]. IADGuard proposes an interpretable defense against structural adversarial attacks on Android malware detection and addresses problem-space attacks that generate executable adversarial software [4]. StratDef uses a moving-target defense strategy to increase adversarial uncertainty through model construction, model selection, and strategic model use [10].

### D. Concept Drift, Novel Families, and Open-Set *Malware Detection*

Malware evolves over time through new families, packed variants, behavioral mimicry, and family-level drift. BenchMFC was introduced to support trustworthy malware family classification under concept drift, including unseen families, packed families, and evolved families [9]. Recent drift work also shows that feature distributions alone may not capture all deployment shifts; stealthy label-inversion drift can occur when test samples retain similar features but change the feature-label relationship [4], [41], [42], [43].

The SHAP-GFR study uses parsed-family LOFO validation to test unseen-family generalization and reports that RF and XGBoost achieved the strongest LOFO performance, while spyware families remained difficult [1]. This motivates a runtime family-novelty mechanism. If a sample is distant from known family profiles, or if its evidence profile resembles previously difficult families, forcing a closed-set prediction may be operationally unsafe.

EGAMA-RC, therefore, treats novelty and family-shift scoring as routing evidence. This does not replace open-set malware detection; rather, it provides a triage mechanism that can flag samples for robust evaluation or analyst review when the known-family evidence is weak.

### *E. Selective Prediction, Calibration, and Operational Triage*

The related works reveals a persistent gap between benchmark performance and operational reliability. Although many malware detection studies [2], [3], [4], [7], [9], [10], [18], [37], [44], [45] report near-perfect accuracy, few provide validated mechanisms for managing uncertain, novel, adversarial, or high-risk cases. Existing work often evaluates explainability, adversarial robustness, drift, and efficiency as separate dimensions, rather than integrating them into a formal policy for abstention, selective prediction, escalation, or analyst review.

This gap motivates EGAMA-RC as an operational extension of SHAP-guided malware detection [1]. Most malware-detection pipelines optimize prediction metrics such as accuracy, precision, recall, F1-score, or ROC-AUC. However, operational triage requires an additional decision: whether the model should act, defer, escalate, or request analyst review. Calibration and selective prediction provide a natural framework for this problem because a calibrated detector can estimate uncertainty more reliably, while a selective classifier can abstain or escalate when risk exceeds a validation-derived threshold.

EGAMA-RC formalizes this idea through evidence-gated triage. Rather than returning a label for every sample through a single model, EGAMA-RC estimates sample-level risk using confidence, uncertainty, model disagreement, explanation stability, perturbation sensitivity, novelty, and latency cost. Predictions are accepted only when risk remains below a validation-calibrated threshold. Otherwise, samples are routed to novelty review, or analyst review. In this way, EGAMA-RC connects classifier evaluation with operational forensic triage by bridging descriptive explanation and prescriptive, risk-calibrated decision-making.

## III. METHODOLOGY

Fig. 1 summarizes the EGAMA-RC study framework and runtime dataflow. The pipeline integrates leakage-controlled preprocessing, dataset-specific feature refinement, model-pool evaluation, sample-level evidence construction, and risk-calibrated routing. Low-risk samples are accepted automatically, while uncertain, high-risk, or novel samples are routed to review, escalation, or novelty-aware handling. The lower audit layer links the framework to clean IID evaluation, robustness testing, open-family analysis, calibration, attribution stability, and deployment-cost assessment.

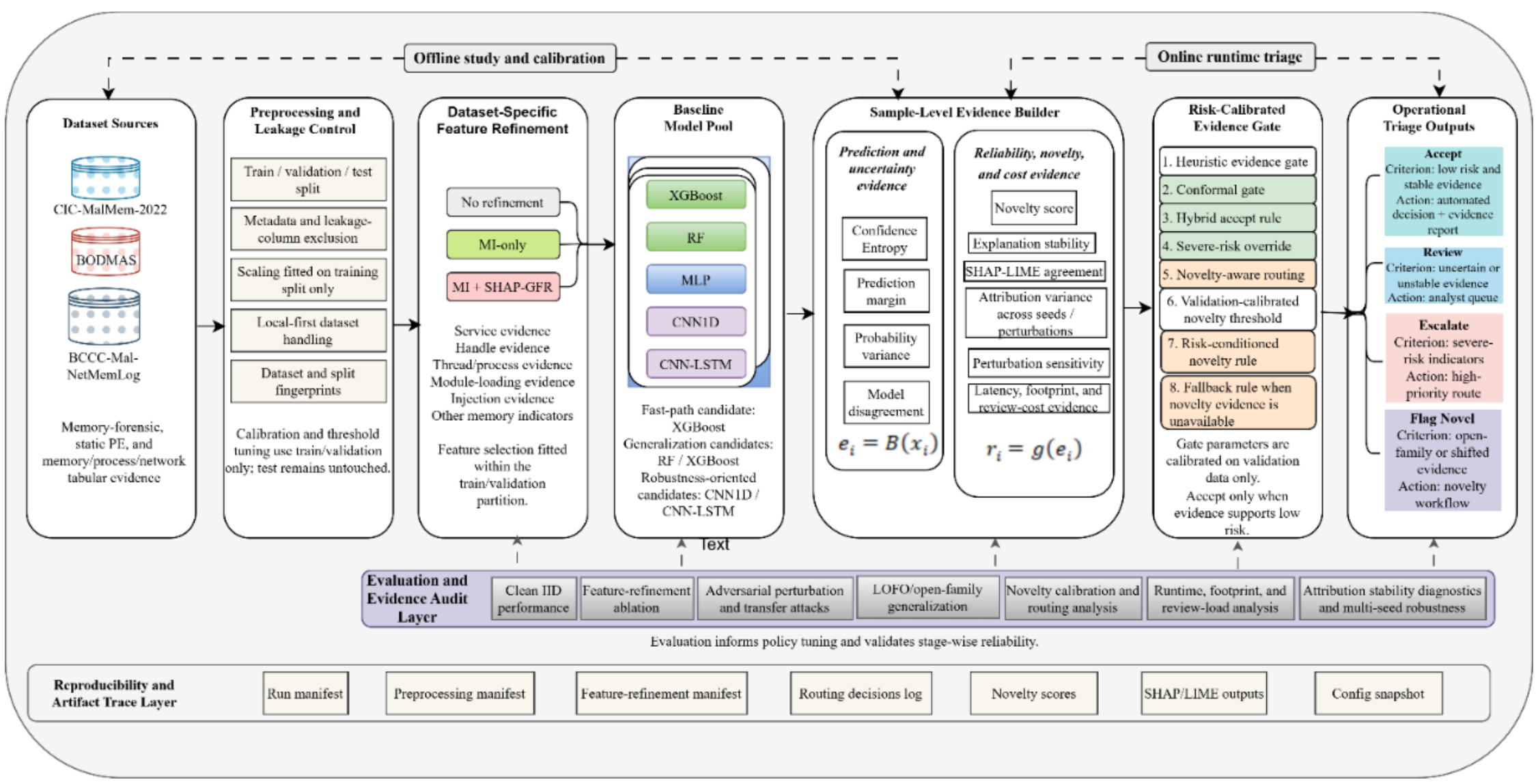

Fig. 1. EGAMA-RC study framework and runtime dataflow.

### A. Problem Definition and Operational Setting

Let each memory-forensic sample be represented by a feature vector defined in Eq. (1).

$$x_i \in \mathbb{R}^d \tag{1}$$

where $d$ denotes the number of retained memory-forensic features. In the original CIC MalMem-2022 SHAP-GFR setting, the refined feature dimension was dataset-specific and depended on the train/validation split and refinement configuration [1]. However, in this study, (d) was treated as a dataset and split-specific quantity rather than a fixed global feature count. Each sample has a binary label $y_i \in \{0,1\}$, where 0 denotes benign and 1 denotes malware. A conventional detector learns a classifier ($f: \mathbb{R}^d \to \{0,1\}$). However, the proposed EGAMA-RC extends this formulation by introducing a triage decision as defined in Eq. (2). Thus, EGAMA-RC is not limited to predicting whether a sample is benign or malicious. It also determines whether the prediction is sufficiently reliable for automated accept, whether specialist models should be invoked, whether the sample should be flagged as family-novel, or whether analyst review is required.

$$a_i \in \{\text{accept,escalate,review,flag-novel}\} \tag{2}$$

### B. Dataset and Feature Evidence Layer

We used three publicly available malware benchmark datasets to assess the performance of the proposed EGAMA-RC framework. The initial evaluation was with the CIC-MalMem-2022 [39], consistent with the previous SHAP-GFR study [1]. This dataset provides memory-resident forensic features extracted from Windows memory snapshots. For CIC MalMem-2022, the feature-refinement pipeline follows the SHAP-GFR procedure from [1], but the retained feature set is learned within the current train/validation split rather than fixed a priori. In the EGAMA-RC framework, feature refinement was evaluated under three modes: no refinement, MI-only refinement, and Mutual information (MI) + SHAP-GFR.

For the CIC-MalMem-2022 [39] baseline evaluation, numeric memory-resident features were normalized using Min–Max scaling fitted only on the training partition. The clean IID protocol followed a stratified 70:15:15 train/validation/test split. For family-level generalization, the parsed-family LOFO protocol removes all samples from the held-out malware family during training and validation and reserves them exclusively for testing.

To strengthen external validity of the EGAMA-RC, we added two more evaluation datasets to support beyond CIC-

MalMem-2022. First, BODMAS[1] [45] was added as a static Windows Portable Executable (PE) feature benchmark. BODMAS provides bodmas.npz feature vectors and aligned metadata, with 134,435 samples, 2,381 features, and binary labels where 0 denotes benign (approximately 57.4%) software and 1 denotes malware (approximately 42.6%). Its metadata were excluded from predictor features to prevent leakage. BODMAS had 581 malware families, and top families such as sfone with 4,729 samples, wacatac with 4,694, upatre with 3,901, and several others above 1,000. BODMAS therefore serves as a cross-modality stress test of the proposed EGAMA-RC framework. It evaluates whether the evidence-gating, feature-refinement, and audit machinery remain valid when the feature space consists of anonymous static PE dimensions rather than named memory-forensic artifacts.

Second, we evaluate EGAMA-RC using the BCCC-Mal-NetMemLog[2] memory CSV resources as a newer memory/process/network tabular benchmark. The current implementation uses category archives and pre-extracted CSV features rather than raw memory dumps or packet captures [46]. Because these resources include very large raw memory and log components, we used CSV resources, process-level logs, metadata, and pre-extracted tabular features rather than raw memory snapshots or PCAP files. The initial BCCC integration focuses on the MemoryCSVs category archives and lightweight merged log files, while heavy components such as the large Sysmon log file and raw memory dumps were excluded in this study. The BCCC dataset had 1,125 benign and 8,498 malware.

Because BODMAS [45] and BCCC [46] datasets differ substantially in class distribution, we supplemented accuracy and weighted scores in this study with macro-F1, balanced accuracy, minority-class recall, worst-class recall, and per-class support. However, test distributions were preserved in their natural form, while class-weighted baselines were evaluated only as controlled training variants. For BODMAS [45] and BCCC [46] dataset, we used each dataset's own native tabular feature space after excluding metadata and leakage columns.

MI preselection was applied using a dataset-scaled candidate-pool policy (see Eq. (3)) before SHAP-GFR. The candidate-pool policy defines how many features are retained after MI preselection and passed to SHAP-GFR. Since CIC MalMem-2022 [39], BCCC MemoryCSVs [46] and BODMAS [45] differ substantially in feature dimensionality and feature semantics, a fixed top-k value would not be proportionally equivalent across datasets.

Therefore, MI preselection was configured in a dataset-scaled manner: smaller memory-forensic feature spaces used smaller candidate pools, while the high-dimensional BODMAS static-PE feature space used a larger capped candidate pool. SHAP-GFR was then applied independently within each dataset's train/validation partition. The automatic MI candidate-pool policy is as defined in Eq. (3), where ($p$) is the number of valid input features after metadata, leakage features and label exclusions. Under this policy, CIC MalMem-2022 resolves to ($k_{MI} = 20$), BODMAS resolves to ($k_{MI} = 256$), and BCCC resolves according to the valid feature count observed.

$$k_{\mathrm{MI}} = \min(\max(20, \lceil 0.20p \rceil), 256) \quad (3)$$

Equation (3) prevents the MI stage from selecting an identical number of features for every dataset. Instead, it defines a dataset-relative candidate feature space for SHAP-GFR, allowing the MI-selected feature count to scale with each dataset's dimensionality while maintaining tractable SHAP computation.

### *C. Baseline Model Pool*

EGAMA-RC inherits the five model families evaluated in the SHAP-GFR study as defined in Eq. (4).

$$\mathcal{F} = \{f_{\mathrm{RF}}, f_{\mathrm{XGB}}, f_{\mathrm{MLP}}, f_{\mathrm{CNN1D}}, f_{\mathrm{CNN\text{-}LSTM}}\} \quad (4)$$

The operational roles are assigned cautiously from validated baseline evidence in [1]. XGBoost was the candidate fast-path model because the current SHAP-GFR study reports comparable clean performance to RF, the lowest median per-sample inference latency, and a small, serialized model size. RF and XGBoost are candidate cross-family generalization models because they achieved the strongest parsed-family LOFO results as reported in [1]. CNN1D and CNN-LSTM were treated as candidate robustness-oriented models because convolutional architectures showed stronger stability under direct gradient-based perturbations in the preceding SHAP-GFR evaluation [1]. LOFO, preprocessing, feature selection, SHAP-GFR refinement, calibration, novelty scoring, and gate thresholds were fitted fold-locally without held-out-family leakage.

The evaluated model pool comprised RF, XGBoost, MLP, CNN1D, and CNN-LSTM using the configurations retained from the original SHAP-GFR study [1]. The MLP used 128- and 64-unit ReLU layers with 0.3 dropout, CNN1D used a 64-filter convolution, max pooling, and a 64-unit dense layer, and CNN-LSTM combined a 64-filter convolution with a 64-unit LSTM and dense layer. XGBoost used 300 estimators, maximum depth 6, learning rate 0.1, and subsample and column-sampling ratios of 0.9, while RF used 200 Gini-based trees with an optional balanced-subsampling variant. Neural models were trained for five epochs with batch size 32 using categorical cross-entropy and Adam. The configurations were retained to evaluate EGAMA-RC as an extension of the same model pool rather than as a hyperparameter-optimized redesign.

### *D. Evidence Signals for Runtime Gating*

For each sample $x_i$, EGAMA-RC constructs an evidence vector from the predicted class-probability vector $p_i \in \mathbb{R}^K$, where $K$ is the number of classes. The implemented evidence vector is defined as in Eq. (5).

$$e_i = [c_i, u_i, m_i, v_i, d_i] \quad (5)$$

where $c_i$ is prediction confidence, $u_i$ is normalized predictive entropy, $m_i$ is the prediction margin, $v_i$ is probability-variance evidence across the model pool, and $d_i$ is model-disagreement evidence.

The confidence score is computed as the maximum predicted class probability as in Eq. (6). The normalized

[1] https://whyisyoung.github.io/BODMAS/

[2] https://www.yorku.ca/research/bccc/ucs-technical/cybersecurity-datasets-cds/large-scale-multisources-malware-analysis-dataset-using-network-traffic-and-memory-bccc-mal-netmem-2025

entropy score is computed as in Eq. (7). The prediction margin is defined as the difference between the largest and second-largest predicted probabilities (see Eq. (8)). Model disagreement was computed from the model-pool predictions. For $M$ models, the modal-frequency disagreement is defined as in Eq. (9).

$$c_i = \max_j p_{ij} \quad (6)$$

$$u_i = -\frac{\sum_{j=1}^{K} p_{ij} \log(p_{ij} + \epsilon)}{\log K} \quad (7)$$

where $\epsilon = 10^{-12}$ is used for numerical stability.

$$m_i = p_{i,(1)} - p_{i,(2)} \quad (8)$$

where $p_{i,(1)}$ and $p_{i,(2)}$ denote the top-ranked and second-ranked class probabilities, respectively.

$$d_i = 1 - \frac{f_{\text{mode},i}}{M} \quad (9)$$

where $f_{\text{mode},i}$ is the number of models predicting the modal class for sample $i$.

Probability variance $v_i$ is computed across model probabilities; for binary classification, it is the variance of the malware-class probability across models, while for multiclass evaluation it is computed from confidence variation across models. Additional evidence sources, including attribution stability, perturbation sensitivity, and runtime overhead, are recorded as diagnostic evidence for robustness analysis, evidence auditing, and deployment-cost interpretation. They were not treated as causal explanations and were not used as direct replacements for the implemented confidence, entropy, margin, variance, and disagreement signals.

### *E. Risk-Calibrated Acceptance and Selective Prediction*

In EGAMA-RC we implement risk-calibrated acceptance through a hybrid of heuristic evidence thresholds and conformal-style calibration. The sample-level risk score used for reporting is defined as in Eq. (10).

$$r_i = 1 - c_i \quad (10)$$

where $c_i$is the prediction confidence.

In the heuristic gate, a sample is accepted only if all active evidence constraints are satisfied. In this study, the default acceptance criteria are defined in Eq. (11).

$$c_i \geq 0.90, u_i \leq 0.50, m_i \geq 0.20, d_i \leq 0.25, v_i \leq 0.05 \quad (11)$$

If any active constraint fails, the sample is routed to review. EGAMA-RC also uses a conformal acceptance condition. On the validation/calibration split, the nonconformity score is defined as in Eq. (12).

$$s_i^{\text{cal}} = 1 - p_i(y_i^{\text{true}}) \quad (12)$$

For $n$ calibration samples and risk level $\alpha$, the conformal cutoff was computed as in Eq. (13).

$$k = \lceil (n+1)(1-\alpha) \rceil, \quad \hat{q} = s_{(k)} \quad (13)$$

At test time, the nonconformity score was (see Eq. (14)):

$$s_i^{\text{test}} = 1 - p_i(y_i^{\text{pred}}) \quad (14)$$

and the conformal gate accepts the sample if $s_i^{\text{test}} \leq \hat{q}$. The default risk level is $\alpha = 0.05$. In hybrid mode, a sample is accepted only when both the heuristic and conformal gates accept, as defined in Eq. (15). This layered policy prevents automatic acceptance when either evidence-threshold criteria or conformal nonconformity evidence indicate elevated uncertainty.

$$a_i^{\text{hybrid}} = a_i^{\text{heuristic}} \wedge a_i^{\text{conformal}} \quad (15)$$

### *F. Adaptive Routing Policy*

EGAMA-RC uses the hybrid gate to assign a binary routing decision and then refines the route with detailed operational labels. The binary route is defined as in Eq. (16). The detailed route is by Eq. (17).

$$r_i^{\text{binary}} \in \{\text{accept}, \text{review}\} \quad (16)$$

$$r_i^{\text{detail}} \in \{\text{accept}, \text{review}, \text{escalate}, \text{flag_novel}\} \quad (17)$$

The hybrid gate first determines whether the sample can be automatically accepted. If the heuristic and conformal conditions both pass, the sample is assigned to accept. Otherwise, it is assigned to a review-level outcome. EGAMA-RC then applies a severe-risk override. A sample is marked for escalation when at least one severe-risk condition is met: $c_i < 0.60 \vee u_i > 0.80 \vee m_i < 0.05 \vee d_i > 0.50 \vee v_i > 0.10$, or when the conformal gate rejects the sample while confidence remains below 0.75 defined in Eq. (18).

$$a_i^{\text{conformal}} \wedge c_i < 0.75. \quad (18)$$

The escalate label, therefore, identifies high-risk cases requiring higher-assurance handling. In the present study, escalation is a routing label; it does not automatically invoke a separate specialist model. Novelty-aware routing is applied after evidence gating. EGAMA-RC builds family profiles over numeric evidence fields, including confidence, entropy, margin, probability variance, and model disagreement. Let $z_i$ denote the evidence-profile vector for the sample $i$, and let $\mu_g$ denote the centroid of the known family $g$ is defined as in Eq. (19).

$$\mu_g = \frac{1}{n_g} \sum_{i:\,\text{family}_i = g} z_i \quad (19)$$

The nearest-family distance is:

$$D_i = \min_g \lVert z_i - \mu_g \rVert_2 \quad (20)$$

The normalized novelty score is:

$$n_i = \frac{D_i}{\max_j D_j + \epsilon} \quad (21)$$

A sample is flagged as open-family risk when its novelty percentile reaches the configured threshold. The default percentile threshold was 95 for this paper. When this condition is met, the detailed route is set to flag_novel, while the collapsed binary route remains review-level for compatibility with accept/review coverage summaries. Thus, the implemented EGAMA-RC route policy is layered rather than a single scalar threshold: evidence extraction, heuristic acceptance, conformal acceptance, hybrid routing, severe-risk escalation, and novelty-aware review are applied in sequence. This design allows the system to preserve automatic acceptance for low-risk samples while routing uncertain, severe-risk, or potentially novel samples away from automatic acceptance. Algorithm 1 outlines the EGAMA-RC adaptive routing steps.

**Algorithm 1: EGAMA-RC Risk-Calibrated Evidence-Gated Routing**

**Input:** sample $x_i$; trained model pool $\mathcal{F}$; fast-path model $f_0$; evidence builder $\mathcal{B}$; heuristic thresholds $\tau_h$; conformal cutoff $\hat{q}$; severe-risk thresholds $\tau_s$; novelty threshold $\tau_{\text{novel}}$.

**Output:** final prediction $\hat{y}_i$; binary route $r_i^{\text{binary}} \in \{\text{accept}, \text{review}\}$; detailed route $r_i^{\text{detail}} \in \{\text{accept}, \text{review}, \text{escalate}, \text{flag_novel}\}$; evidence report $R_i$.

1. Obtain calibrated class-probability outputs from the model pool $\mathcal{F}$.
2. Obtain the fast-path prediction:
$$\hat{y}_i = f_0(x_i)$$
3. Construct the runtime evidence vector:
$$e_i = \mathcal{B}(x_i) = [c_i, u_i, m_i, v_i, d_i]$$
where $c_i$ is confidence, $u_i$ is normalized entropy, $m_i$is margin, $v_i$is probability variance, and $d_i$is model disagreement.
4. Compute the risk reporting score:
$$\rho_i = 1 - c_i.$$
5. Apply the heuristic gate. Set $a_i^{\text{heuristic}} = 1$ if:
$$c_i \geq 0.90, u_i \leq 0.50, m_i \geq 0.20, d_i \leq 0.25, v_i \leq 0.05$$
Otherwise set $a_i^{\text{heuristic}} = 0$.
6. Compute the conformal nonconformity score:
$$s_i^{\text{test}} = 1 - p_i(\hat{y}_i)$$
7. Set $a_i^{\text{conformal}} = 1$ if:
$$s_i^{\text{test}} \leq \hat{q}.$$
Otherwise set $a_i^{\text{conformal}} = 0$.
8. Apply the hybrid gate:
$$a_i^{\text{hybrid}} = a_i^{\text{heuristic}} \wedge a_i^{\text{conformal}}.$$
9. If $a_i^{\text{hybrid}} = 1$, set:
$$r_i^{\text{binary}} = \text{accept}, r_i^{\text{detail}} = \text{accept}.$$
10. Else, set:
$$r_i^{\text{binary}} = \text{review}, r_i^{\text{detail}} = \text{review}.$$
11. Apply the severe-risk rule. If:
$$c_i < 0.60 \vee u_i > 0.80 \vee m_i < 0.05 \vee d_i > 0.50 \vee v_i > 0.10$$
or:
$$\neg a_i^{\text{conformal}} \wedge c_i < 0.75,$$
set:
$$r_i^{\text{detail}} = \text{escalate}.$$
12. Compute novelty evidence using the nearest known-family centroid distance and normalized novelty score $n_i$.
13. If the novelty percentile exceeds $\tau_{\text{novel}}$, set:
$$r_i^{\text{detail}} = \text{flag_novel}, r_i^{\text{binary}} = \text{review}.$$
14. Generate the evidence report:
$$R_i = \{r_i^{\text{binary}}, r_i^{\text{detail}}, \hat{y}_i, \rho_i, c_i, u_i, m_i, v_i, d_i, n_i, \text{route reason}\}$$
15. Record attribution-stability, perturbation-sensitivity, and runtime-overhead evidence as diagnostic artifacts when available.
16. **Return** $\hat{y}_i$, $r_i^{\text{binary}}$, $r_i^{\text{detail}}$, and $R_i$.

### *G. Explanation-Conditioned Evidence Generation*

EGAMA-RC uses SHAP-derived evidence to summarize the forensic basis of prediction. For tree-based models, TreeSHAP was used where supported. For neural models, gradient-based or model-agnostic methods were used only if their stability and computational cost are acceptable. LIME was retained for selected analyst-facing explanations, but SHAP was preferred for consistent attribution over the compact feature set. Explanation evidence was grouped into forensic categories: service evidence, handle evidence, thread/process evidence, module-loading evidence, injection evidence, and other memory indicators. For each sample, explanation stability is measured as defined in Eq. (22).

$$J_k = \frac{| T_k(x_i) \cap T_k(\tilde{x}_i) |}{| T_k(x_i) \cup T_k(\tilde{x}_i) |} \tag{22}$$

where $T_k(x_i)$ is the set of top-$k$ SHAP features for the original sample and $T_k(\tilde{x}_i)$ is the corresponding set after perturbation. EGAMA-RC does not treat SHAP or LIME as causal proof. They are decision-support evidence that were evaluated for stability and fidelity in this paper.

### *H. Threat Model and Adversarial Evaluation Design*

Following NIST AI 100-2e2023 [36], the EGAMA-RC threat model specifies the attacker's objective, capability, knowledge, attack stage, and constraints. The primary objective is to cause malware to be accepted as benign. The attacker may attempt to reduce the probability of malware, decrease the risk score, decrease the novelty score, or create disagreement patterns that exploit the routing policy. Capability levels include:

- black-box attackers observe only labels or limited outputs;
- gray-box attackers know feature categories, model family, or partial score behavior;
- adaptive attacker: knows the gate and attempts to evade both classifier and routing evidence.

The attack stage was primarily test-time evasion. The evaluation includes clean IID testing, multi-seed robustness evaluation, open-family novelty scoring, generic perturbation testing, transfer attacks where supported, attribution-stability diagnostics, and dataset-aware domain perturbations where a valid feature contract is available. Domain perturbation analysis was governed by dataset-specific feature contracts. It was run only when the active feature set satisfies the dataset-specific perturbation contract. When a refinement mode removes one or more required perturbation features, the module is skipped with an explicit manifest rather than forcing an invalid perturbation analysis. An adaptive gate attack is formalized as defined in Eq. (23).

$$\min_{\delta} \; P(y = 1 \mid x_i + \delta) + \lambda_1 r_i(x_i + \delta) + \lambda_2 n_i(x_i + \delta) + \lambda_3 d_i(x_i + \delta) \tag{23}$$

subject to: $x_i + \delta \in \mathcal{C}$, where $\mathcal{C}$ denotes feasible memory-forensic constraints such as non-negative counts, percentile-bounded values, and feature-group plausibility. For differentiable models, attacks are generated by gradient ascent on the classification loss with clipping constraints. Projected Gradient Descent (PGD) update (Eq. (24)), Fast Gradient Sign Method (FGSM) momentum update (Eq. (25)), and MI-FGSM adversarial update (Eq. (26)). Adversarial attack details (gaussian: sigma = 0.02, fgsm: epsilon = 0.02, pgd: epsilon = 0.02, alpha = 0.005, steps = 10, mi_fgsm: epsilon = 0.02, alpha = 0.002, steps = 20, decay = 1.0, transfer: base_attack = mi_fgsm with epsilon/alpha/steps/decay defaults.)

$$x^{(t+1)} = \Pi_{\mathcal{B}_\epsilon(x)}\left(x^t + \alpha \ \text{sign}\left(\nabla_x \mathcal{L}(f(x^t), y)\right)\right) \tag{24}$$

$$m_{t+1} = \mu m_t + \frac{\nabla_x \mathcal{L}(f(x^t), y)}{\|\nabla_x \mathcal{L}(f(x^t), y)\|_1 + \varepsilon} \tag{25}$$

$$x^{(t+1)} = \Pi_{(B_\varepsilon(x))}\left(x^t + \alpha sign\left(m_{(t+1)}\right)\right) \tag{26}$$

where μ denotes the momentum decay factor.

### *I. Latency, Efficiency, and Deployment Protocol*

The latency and efficiency protocol focused on deployability evidence for EGAMA-RC, including fast-path inference latency, p50/p95 per-sample latency, serialized model size, neural-model parameter count, analyst-review rate, escalation rate, novelty burden, and safety–throughput trade-offs. This study did not separately decompose gate-overhead latency, novelty-scoring overhead, explanation-generation overhead, or route-specific review/escalation p50/p95 latency.

Therefore, complete decision-path observability is treated as a deployment limitation and future extension rather than as a fully measured component in the present study. Our previous SHAP-GFR study [1] already reports model-specific efficiency, including XGBoost's low median per-sample latency and compact model size. EGAMA-RC extends this to route-specific latency, because a triage system may incur different computational and reporting costs depending on whether a sample is accepted, escalated, reviewed, or flagged as novel.

### *J. Standard Classification Evaluation Metrics*

For standard classification performance, we report precision, recall, F1-score, balanced accuracy, false-positive rate (FPR), false-negative rate (FNR), and area under the precision–recall curve (AUPRC). Precision, recall, and F1-score are defined as Eqs. (27), (28), and (29), respectively.

$$\text{Precision} = \frac{TP}{TP + FP} \tag{27}$$

$$\text{Recall} = \frac{TP}{TP + FN} \tag{28}$$

$$\text{F1} - \text{score} = 2 \cdot \frac{\text{Precision} \cdot \text{Recall}}{\text{Precision} + \text{Recall}} \tag{29}$$

For EGAMA-RC routing, accepted coverage measures the fraction of samples assigned to the automatic-acceptance route, defined as in Eqs. (30)-(34).

$$\text{Accepted Coverage} = \frac{N_{\text{accept}}}{N} \tag{30}$$

$$\text{Review Rate} = \frac{N_{\text{review}}}{N} \tag{31}$$

$$\text{Escalation Rate} = \frac{N_{\text{escalate}}}{N} \tag{32}$$

$$\text{Novelty Flag Rate} = \frac{N_{\text{flag_novel}}}{N} \tag{33}$$

The false-accept malware rate (FAMR) measures the fraction of true malware samples that are incorrectly accepted as benign (see Eq. (34))

$$\text{FAMR} = \frac{\sum_{i=1}^{N} \mathbb{1}\,[y_i = 1 \wedge \hat{y}_i = 0 \wedge r_i = \text{accept}]}{\sum_{i=1}^{N} \mathbb{1}\,[y_i = 1]} \tag{34}$$

### *K. Reproducibility and Implementation Details*

The EGAMA-RC implementation was organized as an experimental pipeline with dedicated configuration files, artifact directories, tests, runtime logs, evidence audits, and multi-seed aggregation outputs. All feature-refinement artifacts record the selected mode, MI candidate-pool policy, scorer requested and used, fallback status, leakage guard, dataset fingerprint, split fingerprint, and selected feature set. This prevents a SHAP-GFR artifact from being reused as MI-only evidence or a dataset-specific feature subset from being transferred across incompatible datasets. All final experiments were repeated across five random seeds (13, 21, 42, 87, and 101) and the corresponding split, feature-refinement, model, and evaluation seed values were recorded in the run manifests for reproducibility. All experiments were conducted on a Windows platform (Lenovo, 12th Gen Intel (R) Core (TM) i7-1260P (16 CPUs) ~2.1GHz, 32GB RAM). The experimental environment used Python 3.11 with pandas 2.3.3, NumPy 2.2.6, scikit-learn 1.7.2, XGBoost 3.2.0, TensorFlow 2.21.0, PyTorch 2.10.0+cu128, SHAP 0.49.1, LIME 0.2.0.1, matplotlib 3.10.8, seaborn 0.13.2, and PyYAML 6.0.3; CUDA 12.8 was available for GPU-supported runs.

## IV. RESULTS AND DISCUSSIONS

### *A. Multi-Dataset Feature Refinement and Baseline Model Performance*

We first evaluate the clean IID baseline to establish whether the feature-refinement layer and model pool are sufficiently reliable before EGAMA-RC routing is applied. The three datasets differ substantially in feature dimensionality and class balance. Table I summarizes the feature-refinement behavior across the three study datasets and five random seeds. With no refinement, the full valid feature space is retained. The MI-only setting consistently retains 20 features for CIC-MalMem-2022 and BCCC, and 256 features for BODMAS. SHAP-GFR further adapts the retained feature set by dataset and seed: CIC-MalMem-2022 retains 15–20 features, BCCC retains 17–20 features, and BODMAS retains 159–256 features. This confirms that the refinement stage is not applying a fixed feature count across datasets but instead adjusts to the dimensionality and validation behavior of each feature space.

Table I. Dataset characteristics and feature-refinement outcomes across the three EGAMA-RC

| Dataset | Valid input features | MI-only retained features | SHAP-GFR retained features across seeds |
|---|---|---|---|
| CIC-MalMem-2022 | 53 | 20 | 15, 15, 15, 20, 15 |
| BODMAS | 2,381 | 256 | 256, 218, 159, 256, 218 |
| BCCC | 94 | 20 | 20, 17, 20, 17, 17 |

Fig. 2 shows the clean IID results. It reveals that CIC-MalMem-2022 and BODMAS remain highly separable across all refinement settings. On CIC-MalMem-2022, all three refinement modes achieve near-ceiling performance, with accuracy, macro-F1, balanced accuracy, and AUPRC approximately 0.9999–1.00. BODMAS also remains strong across all settings, with the full-feature baseline marginally highest on aggregate clean metrics, while MI-only and SHAP-GFR remain close. BCCC was seen to be the most challenging dataset because of its severe imbalance. On BCCC, feature refinement improves class-sensitive behavior: SHAP-GFR increases macro-F1 from 0.5432 to 0.5694 and minority-class recall from 0.0828 to 0.1254 relative to no refinement, although AUPRC remains slightly higher without refinement.

Finally, we compare XGBoost and MLP under clean IID evaluation before routing (see Table II). We observed that the XGBoost was consistently strongest across all three datasets on accuracy, macro-F1, balanced accuracy, minority-class recall, and AUPRC. Its AUPRC values were 1.00 on CIC-MalMem-2022, 0.9739 on BCCC, and 0.9998 on BODMAS, exceeding the corresponding MLP values of 0.9999, 0.9324, and 0.9683. This supports the use of XGBoost as the clean-data fast-path model in EGAMA-RC, while later routing stages are needed to handle uncertainty, novelty, adversarial sensitivity, and review cost.

Thus, these baseline results show that the model pool is competent before routing is introduced. CIC-MalMem-2022 and BODMAS exhibit strong clean separability, while BCCC exposes the expected difficulty of class imbalance and minority-class detection. Feature refinement is therefore most useful as a stabilizing and class-sensitive mechanism on the harder dataset, and XGBoost provides the strongest clean IID baseline across the evaluated model-pool metrics.

Table II. Clean IID model-pool comparison between XGBoost and MLP using AUPRC

| Dataset | XGBoost AUPRC | MLP AUPRC | Stronger model |
|---|---|---|---|
| CIC-MalMem-2022 | 1.00 | 0.9999 | XGBoost |
| BCCC | 0.9739 | 0.9324 | XGBoost |
| BODMAS | 0.9998 | 0.9683 | XGBoost |

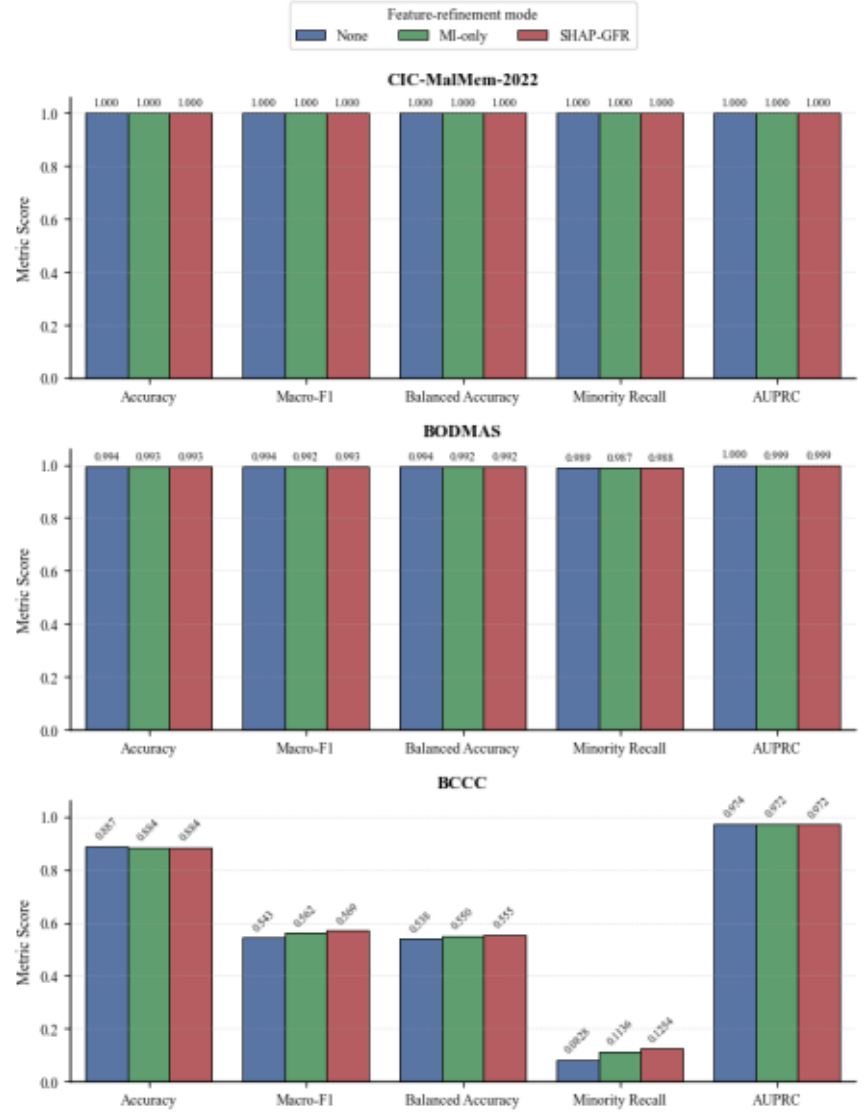

Fig. 2. Clean IID baseline performance under no refinement, MI-only refinement, and SHAP-GFR across CIC-MalMem-2022, BODMAS, and BCCC.

*B. Risk-Calibrated Routing and Safe-Acceptance Behavior*

We next evaluate whether EGAMA-RC can safely automate low-risk cases before novelty-aware routing is applied. This analysis focuses on the evidence-gated routing layer and compares three gate configurations: heuristic-only, conformal-only, and hybrid. Table III presents the results. On the pooled test set of 152,000 samples, the hybrid gate accepted 93.12% of samples while maintaining 99.864% accepted-set accuracy, a 0.136% false-accept rate, and a 0.106% false-accept malware rate. The corresponding review rate was 6.88%, while severe-risk escalation remained active for 2.35% of samples. Relative to the uncalibrated baseline policy, accept coverage increased from 64.65% to 93.12%, review load decreased from 35.35% to 6.88%, and the false-accept rate decreased from 0.195% to 0.136%.

Table III. Overall evidence-gate comparison before novelty-aware routing.

| Gate mode | AC (%) | RR (%) | ACC (%) | FAR (%) | FAMR (%) | ER (%) |
|---|---|---|---|---|---|---|
| Heuristic | 93.62 | 6.38 | 99.863 | 0.137 | 0.108 | 2.35 |
| Conformal | 96.31 | 3.69 | 99.360 | 0.640 | 0.280 | 2.35 |
| Hybrid | 93.12 | 6.88 | 99.864 | 0.136 | 0.106 | 2.35 |

Note: AC = Accept coverage, RR = Review rate, ACC = Accepted accuracy, FAR = False-accept rate, FAMR = False-accept malware rate, ER = Escalation rate

Fig. 3 shows a clear safety-throughput frontier. The conformal-only gate achieves the highest accept coverage, accepting 96.31% of samples and reducing review to 3.69%. However, this gain in coverage comes with a considerably higher false-accept rate of 0.640% and a higher false-accept malware rate of 0.280%. By contrast, the heuristic and hybrid gates are more conservative. The hybrid gate retains high automation while achieving the lowest false-accept rate and false-accept malware rate among the three gate modes. This makes hybrid gating the preferred operating point for safety-critical malware triage, where unsafe automated acceptance is more costly than additional review.

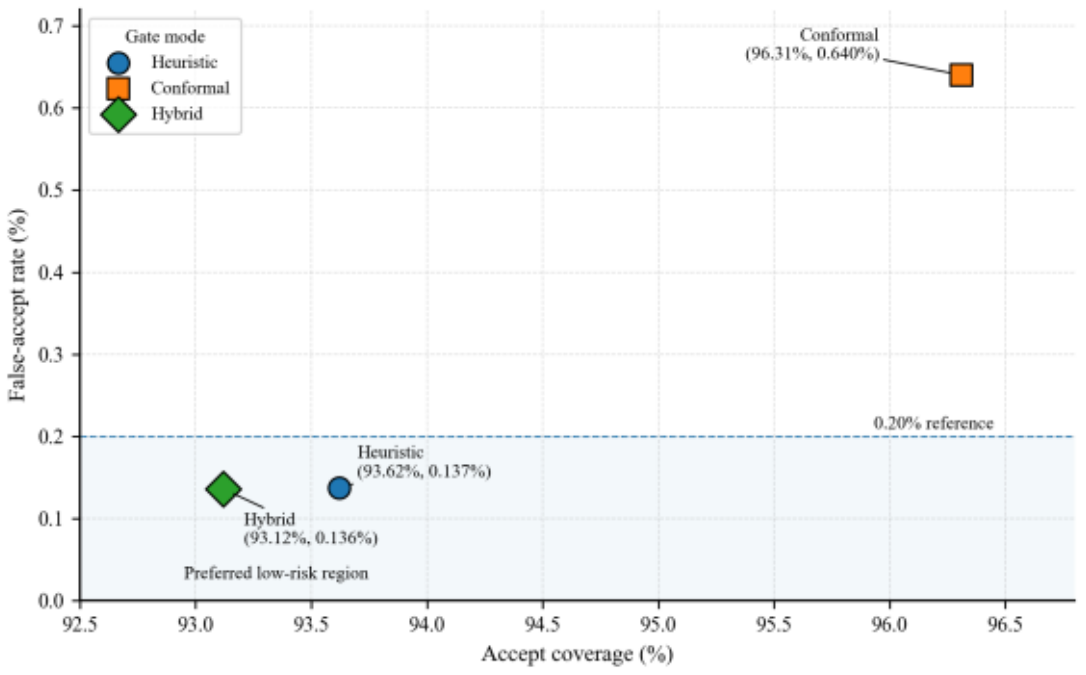

Fig. 3. Safety-throughput trade-off across EGAMA-RC gate configurations. Plot accept coverage on the x-axis and false-accept rate on the y-axis for heuristic, conformal, and hybrid gates.

Dataset-level routing further confirms that the gate adapts to workload difficulty (see Table IV). CIC-MalMem-2022 and BODMAS support high-confidence automation, with hybrid accepted accuracies of 100% and 99.92%, respectively. BCCC was the hardest dataset: the hybrid gate accepts only 60.04% of samples and routes 39.96% to review. This restriction was appropriate because conformal-only routing accepts 100% of BCCC samples but reduces accepted accuracy to 88.42% and increases the false-accept rate to 11.58%. Thus, the hybrid gate does not simply maximize throughput; it suppresses automatic acceptance when the evidence profile indicates higher risk.

Table IV. Dataset-level hybrid routing performance before novelty-aware routing.

| Dataset | AC (%) | RR (%) | ACC (%) | FAR (%) | FAMR (%) | ER (%) |
|---|---|---|---|---|---|---|
| BCCC | 60.04 | 39.96 | 97.21 | 2.79 | 0.00 | 18.01 |
| BODMAS | 93.34 | 6.66 | 99.92% | 0.0765 | 0.1738 | 2.23% |

| CIC-MalMem-2022 | 98.04 | 1.96 | 100.00 | 0.00 | 0.00 | 0.0637 |
|---|---|---|---|---|---|---|

Note: AC = Accept coverage, RR = Review rate, ACC = Accepted accuracy, FAR = False-accept rate, FAMR = False-accept malware rate, ER = Escalation rate

Severe-risk escalation was dominated by high predictive entropy, which triggered on 2.3546% of all samples (see Table V). Confidence and margin triggers were less frequent, at 0.6914% and 0.1559%, respectively. Model-disagreement and probability-variance triggers did not activate at the configured thresholds. The conformal-reject-plus-low-confidence condition occurred less often under conformal and hybrid gating than under heuristic-only gating, decreasing from 2.1987% to 1.3914%.

Table V. Severe-risk trigger breakdown across evidence-gate configurations.

| Severe-risk trigger | Heuristic (%) | Conformal (%) | Hybrid (%) |
|---|---|---|---|
| Entropy above threshold | 2.355 | 2.355 | 2.355 |
| Confidence below threshold | 0.691 | 0.691 | 0.691 |
| Margin below threshold | 0.156 | 0.156 | 0.156 |
| Conformal reject + low confidence | 2.199 | 1.391 | 1.391 |
| Model disagreement above threshold | 0.00 | 0.00 | 0.00 |
| Probability variance above threshold | 0.00 | 0.00 | 0.00 |

Largely, the evidence-gated results show that EGAMA-RC can safely automate low-risk malware-triage decisions. The hybrid gate accepts more than 93% of pooled samples while maintaining a very low unsafe-accept profile. It also preserves review and escalation pathways for harder cases, especially BCCC, where a less restrictive conformal-only gate would produce substantially higher false accepts. These findings support the central EGAMA-RC claim: reliable malware triage requires selective, evidence-gated automation rather than unconditional prediction coverage.

### *C. Novelty-Aware Routing and Calibration Effects*

Table VI shows the effect of novelty calibration on EGAMA-RC routing performance. The selected policy, corresponding to a novelty quantile of 0.80, conformal alpha of 0.01, and risk-conditioned novelty enabled. We observed that, relative to the uncalibrated baseline, the selected policy increased accept coverage from 64.65% to 93.12%, while accepted accuracy improved slightly from 99.80% to 99.86% and the false-accept rate decreased from 0.20% to 0.14%. The selected policy also reduced review volume from 35.35% to 6.88%, indicating that calibrated novelty routing recovered substantial automation capacity without weakening the safe-acceptance profile. The open-family rate decreased from 32.60% to 11.54%, while the escalation rate remained unchanged at 2.35%. The selected configuration produced zero conformal rejections, indicating that the final routing behavior was driven primarily by calibrated novelty and risk-conditioned evidence rather than conformal rejection.

Thus, calibration-guided EGAMA-RC routing improved acceptance efficiency without degrading safety, increasing accept coverage by 28.47 percentage points, while maintaining near-constant accepted accuracy and reducing false-accept rate.

Table VI: Effect of novelty calibration on EGAMA-RC routing performance.

| Metric | Uncalibrated baseline (%) | Selected calibrated policy (%) |
|---|---|---|
| Accept coverage | 64.65 | 93.12 |
| Review rate | 35.35 | 6.88 |
| Accepted accuracy | 99.80 | 99.86 |
| False-accept rate | 0.20 | 0.14 |
| Open-family rate | 32.60 | 11.54 |
| Escalation rate | 2.35 | 2.35 |
| Conformal reject rate | 0.00 | 0.00 |

### *D. Robustness, Open-Family Generalization, and Explanation Stability*

We evaluated EGAMA-RC under perturbation, family shift, novelty exposure, and explanation instability. The results show stronger cross-family generalization than adversarial robustness. Under FGSM, MLP accuracy fell to 0.3804 with ΔF1 = -0.8530 and 61.96% attack success, while FGSM-transfer reduced XGBoost accuracy to 0.4829 with 51.71% attack success. These findings show that the classifiers are not uniformly attack-invariant and support the use of evidence-gated deferral and escalation rather than unconditional prediction.

LOFO results were substantially stronger. RF and XGBoost achieved mean F1-scores of 0.9988 and 0.9987, respectively, with worst-family F1-scores of 0.9876 and 0.9874 on Transponder. Neural models were more sensitive to family shift, reaching F1-scores of 0.8741, 0.8714, and 0.8552 for MLP, CNN1D, and CNN-LSTM on the hardest held-out family, TIBS. This indicates that tree ensembles provide more stable family-level generalization.

Novelty results also demonstrated the need for dataset-conditioned calibration. At the fixed 95th-percentile threshold, mean open-family rates were 0.0515 for BODMAS and 0.0565 for BCCC, whereas all 8,790 CIC-MalMem-2022 test samples were flagged. This review-all outcome reflects an over-restrictive fixed threshold rather than universal novelty and motivates the calibrated novelty policy in Section IV-C.

Domain-aware perturbation was evaluated only on CIC-MalMem-2022 because valid feature contracts were unavailable for BODMAS and BCCC. Across three seeds, perturbations of magnitude 0.10 produced 8,790 test outputs and affected 13 features spanning handle-density, service, thread-process, injection, and module-loading groups. Under MLP FGSM, attribution rankings remained moderately stable, with Spearman correlation of 0.7186, Kendall correlation of 0.6779, top-5 overlap of 0.95, top-10 overlap of 0.9875, cosine similarity of 0.7456, and a 12.5% prediction-flip rate.

Thus, EGAMA-RC supports operational robustness by using uncertainty, disagreement, novelty, and perturbation evidence to determine when automated acceptance is appropriate. It does not claim worst-case resistance to all gradient-based or transfer attacks; rather, its contribution lies in risk-calibrated routing that defers or escalates unreliable cases.

*E. Runtime Overhead and Deployment Trade-Offs*

We finally evaluated the deployment implications of EGAMA-RC in terms of inference latency, model footprint, analyst-review load, and safety-throughput trade-offs (see Table VII). The measured fast-path inference latency is highly model-dependent. XGBoost provides the lowest per-sample latency, with p50/p95 latency of 0.0054/0.0059 ms, followed by RF at 0.1839/0.3715 ms. Among neural models, CNN-LSTM records 0.4711/0.7613 ms, MLP records 0.6835/0.8598 ms, and CNN1D records 0.7299/0.9406 ms. These results support XGBoost as the most efficient fast-path model, while RF remains a practical low-latency non-neural alternative.

Table VII. Fast-path inference latency and model footprint

| Model | p50 latency (ms) | p95 latency (ms) | Model size (MB) | Parameters |
|---|---|---|---|---|
| XGBoost | 0.0054 | 0.0059 | 0.312 | N/A |
| RF | 0.1839 | 0.3715 | 1.341 | N/A |
| CNN-LSTM | 0.4711 | 0.7613 | 0.470 | 37,570 |
| MLP | 0.6835 | 0.8598 | 0.147 | 10,178 |
| CNN1D | 0.7299 | 0.9406 | 0.320 | 25,026 |

We observed small model footprints across most models. The MLP has the smallest footprint at 0.147 MB, followed by XGBoost at 0.312 MB, CNN1D at 0.320 MB, and CNN-LSTM at 0.470 MB. RF is larger at 1.341 MB but remains modest in absolute size. Parameter counts are available for neural models only: MLP has 10,178 parameters, CNN1D has 25,026, and CNN-LSTM has 37,570. Since tree ensembles do not have trainable parameters in the same sense as neural networks, RF and XGBoost were compared using serialized model footprints rather than parameter count.

Deployment cost is not determined by latency alone; it is also shaped by the review burden induced by the gate. On the pooled evaluation set of 152,000 samples, conformal-only gating routes 5,607 samples to review, corresponding to a review rate of 3.69% (see Table VIII). Heuristic gating routes 9,692 samples to review, corresponding to 6.38%, while hybrid gating routes 10,465 samples to review, corresponding to 6.88%. Thus, conformal gating minimizes analyst workload, but this comes with a higher false-accept rate.

Table VIII. Analyst-review load and safety-throughput trade-off by gate mode

| Gate mode | Accept coverage | Review count | Review rate | False-accept rate |
|---|---|---|---|---|
| Conformal | 96.31% | 5,607 | 3.69% | 0.640% |
| Heuristic | 93.62% | 9,692 | 6.38% | 0.137% |
| Hybrid | 93.12% | 10,465 | 6.88% | 0.136% |

The safety–throughput trade-off is clear. Conformal gating achieves the highest accept coverage at 96.31% but increases the false-accept rate to 0.640%. Heuristic and hybrid gating accept 93.62% and 93.12% of samples, respectively, while reducing the false-accept rate to approximately 0.14%. For safety-critical malware triage, the lower unsafe-accept rate justifies the modest increase in review load. Novelty pressure also varies by dataset. Under the fixed threshold-95 policy, mean open-family rates are 5.15% for BODMAS and 5.65% for BCCC, whereas all CIC-MalMem-2022 samples are flagged. This reflects an over-restrictive fixed threshold rather than universal novelty and supports the use of dataset-calibrated novelty policies. The deployment analysis is limited to fast-path inference latency, model footprint, review load, novelty burden, and safety–throughput behavior. Gate, novelty-scoring, explanation-generation, and route-specific latencies were not measured separately, so end-to-end decision latency may exceed base inference time.

In practice, the operating policy can be matched to organizational risk tolerance. Heuristic routing offers low false-accept risk with moderate review load, calibrated novelty is suitable when unknown-family exposure is important, and hybrid routing is preferable when safety outweighs throughput. EGAMA-RC provides a lightweight and configurable triage framework that exchanges modest review overhead for substantially safer automated acceptance.

*F. Comparison with Related Work*

Representative malware-detection studies address important but largely separate deployment concerns. Prior memory-forensic and dynamic-analysis approaches report strong performance on obfuscated, unseen, or previously unseen malware [13], [18], while BenchMFC evaluates family classification under unseen-family, packed-family, and temporal drift [9]. Other studies focus on adversarial defense: StratDef applies moving-target strategies across several attack methods [10], MalProtect uses stateful threat indicators to reduce adversarial query evasion [2], and IADGuard addresses structural problem-space attacks against Android malware detectors [4]. These works advance robustness, drift handling, explanation, or runtime monitoring, but they do not jointly implement calibrated acceptance, abstention, escalation, novelty-aware handling, and analyst review within a unified malware-triage policy.

Direct numerical comparison is inappropriate because the studies use different datasets, feature representations, splits, attack models, and evaluation objectives. The distinction of EGAMA-RC is therefore methodological rather than a claim of universal performance superiority. EGAMA-RC combines predictive uncertainty, model disagreement, novelty evidence, explanation diagnostics, and risk calibration to determine whether each prediction should be accepted automatically or routed to review, escalation, or novelty-aware handling. In the pooled evaluation, the selected hybrid gate accepted 93.12% of samples with 99.864% accepted-set accuracy, a 0.136% false-accept rate, and a 0.106% false-accept malware rate. These results position EGAMA-RC as an operational risk-control layer rather than a worst-case adversarially robust classifier.

## V. DISCUSSION

The findings support a central claim: reliable malware triage is not only a modeling problem but also a decision-governance problem. Predictive performance becomes operationally useful only when coupled with routing logic that separates confident acceptance from uncertain, high-risk, or potentially novel cases. EGAMA-RC therefore derives its value not from a single classifier, but from the controlled allocation of trust across acceptance, review, escalation, and novelty-aware pathways.The results also show that robustness is not a single property. Generic perturbations, transfer attacks, domain-aware stress tests, and open-family evaluation expose different failure modes, and no model family dominates across all conditions. Tree ensembles provide strong clean and LOFO performance, whereas neural models respond differently under gradient-based perturbations. EGAMA-RC addresses this variation by converting uncertainty and risk evidence into routing decisions. The practical objective is therefore not complete attack resistance, but safer system behavior when model resistance is imperfect.

Novelty-aware routing is particularly important because unseen families and distribution shifts are expected in malware operations. A calibrated novelty mechanism allows unfamiliar evidence to trigger review rather than forcing a closed-set prediction. This improves containment discipline, although it may increase analyst workload when novelty exposure is high. Explanation stability provides an additional reliability signal by showing whether the forensic basis of a prediction remains coherent under perturbation. EGAMA-RC therefore treats interpretability as diagnostic evidence rather than as a post-hoc visualization alone. Deployment results further clarify the safety-throughput trade-off. Higher accept coverage reduces review load but may increase unsafe acceptance, whereas stricter policies shift more samples to review or escalation. The appropriate operating point therefore depends on organizational risk tolerance. Heuristic or hybrid routing is suitable when low false acceptance is prioritized, while calibrated novelty routing is preferable when unknown-family exposure is a major concern.

Several limitations bound these findings. The current artifacts do not separately measure gate, novelty-scoring, explanation-generation, or route-specific latency. The robustness results are limited to the evaluated datasets, attacks, and perturbation contracts, and stronger adaptive adversaries may expose additional weaknesses. Novelty behavior is also dataset dependent, model-footprint metadata are not equally available across model families, and analyst impact is represented through review-load proxies rather than direct user studies. Despite these limitations, the study demonstrates a reproducible method for connecting model evaluation with operational decision policy.

## VI. CONCLUSION

This study presents EGAMA-RC as a risk-calibrated malware-triage framework that extends detection beyond unconditional classification. The framework uses confidence, uncertainty, disagreement, novelty, and severe-risk evidence to determine whether a sample should be accepted automatically, reviewed, escalated, or flagged as novel. Across the evaluated datasets, hybrid routing provides a strong safety-throughput operating point, while calibrated novelty reduces over-restrictive review behavior without weakening the safe-acceptance profile. The results indicate that the main operational value of malware detection lies not only in predictive accuracy, but also in controlling when automated decisions should be trusted. Future work will examine full decision-path latency, adaptive attacks against the complete routing stack, temporal and cross-organization validation, improved novelty calibration, analyst-in-the-loop evaluation, and cost-aware policy optimization. These extensions will further clarify how evidence-gated routing can support deployment in operational malware-analysis environments.

## REFERENCES

[1] I. K. Nti and O. Nyarko-Boateng, "SHAP-Guided Feature Refinement for Robust and Interpretable Malware Detection in Memory Forensics," Nov. 14, 2025. doi: 10.21203/rs.3.rs-7871553/v1.

[2] A. Rashid and J. Such, "MalProtect: Stateful Defense Against Adversarial Query Attacks in ML-Based Malware Detection," *IEEE Transactions on Information Forensics and Security*, vol. 18, pp. 4361–4376, 2023, doi: 10.1109/TIFS.2023.3293959.

[3] X. Deng, H. Tang, X. Pei, D. Li, and K. Xue, "MDHE: A Malware Detection System Based on Trust Hybrid User-Edge Evaluation in IoT Network," *IEEE Transactions on Information Forensics and Security*, vol. 18, pp. 5950–5963, 2023, doi: 10.1109/TIFS.2023.3318947.

[4] W. Wei *et al.*, "Interpretable Defense Against Structural Adversarial Attacks on Android Malware Detection," *IEEE Transactions on Information Forensics and Security*, vol. 20, pp. 13296–13311, 2025, doi: 10.1109/TIFS.2025.3639962.

[5] M. Saqib, S. Mahdavifar, B. C. M. Fung, and P. Charland, "A Comprehensive Analysis of Explainable AI for Malware Hunting," *ACM Comput. Surv.*, vol. 56, no. 12, Oct. 2024, doi: 10.1145/3677374.

[6] H. Manthena, S. Shajarian, J. C. Kimmell, M. Abdelsalam, S. Khorsandroo, and M. Gupta, "Explainable Artificial Intelligence (XAI) for Malware Analysis: A Survey of Techniques, Applications, and Open Challenges," *IEEE Access*, vol. 13, pp. 61611–61640, 2025, doi: 10.1109/ACCESS.2025.3555926.

[7] J. Jeon, B. Jeong, S. Baek, and Y. S. Jeong, "Hybrid Malware Detection Based on Bi-LSTM and SPP-Net for Smart IoT," *IEEE Trans. Industr. Inform.*, vol. 18, no. 7, pp. 4830–4837, Jul. 2022, doi: 10.1109/TII.2021.3119778.

[8] E. Baghirov, "A comprehensive investigation into robust malware detection with explainable AI," *Cyber Security and Applications*, vol. 3, Dec. 2025, doi: 10.1016/j.csa.2024.100072.

[9] Y. Jiang, G. Li, S. Li, and Y. Guo, "BenchMFC: A benchmark dataset for trustworthy malware family classification under concept drift," *Comput. Secur.*, vol. 139, Apr. 2024, doi: 10.1016/j.cose.2024.103706.

[10] A. Rashid and J. Such, "StratDef: Strategic defense against adversarial attacks in ML-based malware detection," *Comput. Secur.*, vol. 134, Nov. 2023, doi: 10.1016/j.cose.2023.103459.

[11] S. Kharnotia, B. Arora, and R. Kour, "Feature-driven static analysis for learning-based android malware detection: A review," *ICT Express*, vol. 12, no. 1, pp. 186–208, Feb. 2026, doi: 10.1016/j.icte.2026.01.005.

[12] A. Al Saaidah *et al.*, "Enhancing Malware Detection Performance: Leveraging K-Nearest Neighbors with Firefly Optimization Algorithm," *Multimed. Tools Appl.*, vol. 84, no. 12, pp. 10071–10094, Mar. 2024, doi: 10.1007/s11042-024-18914-5.

[13] O. A. Madamidola, F. Ngobigha, and A. Ez-zizi, "Detecting new obfuscated malware variants: A lightweight and interpretable machine learning approach," *Intelligent Systems with Applications*, vol. 25, Mar. 2025, doi: 10.1016/j.iswa.2024.200472.

[14] S. F. Bilal, S. Bashir, F. H. Khan, H. Rasheed, and S. and K. F. H. and R. H. Bilal Syed Fakhar and Bashir, "Malwares Detection for Android and Windows System by Using Machine Learning and Data

Mining," F. and C. A. Bajwa Imran Sarwar and Kamareddine, Ed., Singapore: Springer Singapore, 2019, pp. 485–495. doi: 10.1007/978-981-13-6052-7_42.

[15] Q. Qiang, Y. Chen, Y. Hu, T. Zang, and M. Cheng, *Cost-Effective Malware Classification*, vol. 3. Springer Nature Switzerland, 2023. doi: 10.1007/978-3-031-25538-0.

[16] A. Odeh, A. A. Taleb, T. Alhajahjeh, and F. Navarro, "Advanced memory forensics for malware classification with deep learning algorithms," *Cluster Comput.*, vol. 28, no. 6, p. 353, Oct. 2025, doi: 10.1007/s10586-025-05104-7.

[17] H. Mohammadian, G. Higgins, S. Ansong, R. Razavi-Far, and A. A. Ghorbani, "Explainable malware detection through integrated graph reduction and learning techniques," *Big Data Research*, vol. 41, p. 100555, Aug. 2025, doi: 10.1016/j.bdr.2025.100555.

[18] R. K. Koppanati, M. Santra, and S. Kumar Peddoju, "D24D: Dynamic Deep 4-Dimensional Analysis for Malware Detection," *IEEE Transactions on Information Forensics and Security*, vol. 20, pp. 2083–2095, 2025, doi: 10.1109/TIFS.2025.3531230.

[19] K. Liu, S. Xu, G. Xu, M. Zhang, D. Sun, and H. Liu, "A Review of Android Malware Detection Approaches Based on Machine Learning," *IEEE Access*, vol. 8, pp. 124579–124607, 2020, doi: 10.1109/ACCESS.2020.3006143.

[20] J.-Y. Kim and S.-B. Cho, "Obfuscated Malware Detection Using Deep Generative Model based on Global/Local Features," *Comput. Secur.*, vol. 112, p. 102501, Jan. 2022, doi: 10.1016/j.cose.2021.102501.

[21] M. S. Nawaz, P. Fournier-Viger, M. Z. Nawaz, G. Chen, and Y. Wu, "MalSPM: Metamorphic malware behavior analysis and classification using sequential pattern mining," *Comput. Secur.*, vol. 118, p. 102741, Jul. 2022, doi: 10.1016/j.cose.2022.102741.

[22] V. Ravi and M. Alazab, "Attention-based convolutional neural network deep learning approach for robust malware classification," *Comput. Intell.*, vol. 39, no. 1, pp. 145–168, Feb. 2023, doi: 10.1111/coin.12551.

[23] M. Torres, R. Alvarez, and M. Cazorla, "A Malware Detection Approach Based on Feature Engineering and Behavior Analysis," *IEEE Access*, vol. 11, pp. 105355–105367, 2023, doi: 10.1109/ACCESS.2023.3319093.

[24] M. M. Alani, A. Mashatan, and A. Miri, "XMal: A lightweight memory-based explainable obfuscated-malware detector," *Comput. Secur.*, vol. 133, Oct. 2023, doi: 10.1016/j.cose.2023.103409.

[25] M. Botacin, F. Ceschin, R. Sun, D. Oliveira, and A. Grégio, "Challenges and pitfalls in malware research," *Comput. Secur.*, vol. 106, p. 102287, 2021, doi: 10.1016/j.cose.2021.102287.

[26] M. Stamp, M. Alazab, and A. Shalaginov, *Malware analysis using artificial intelligence and deep learning*. Springer International Publishing, 2020. doi: 10.1007/978-3-030-62582-5.

[27] R. Vinayakumar, M. Alazab, K. P. Soman, P. Poornachandran, and S. Venkatraman, "Robust Intelligent Malware Detection Using Deep Learning," *IEEE Access*, vol. 7, pp. 46717–46738, 2019, doi: 10.1109/ACCESS.2019.2906934.

[28] D. Vu, T. Nguyen, T. V. Nguyen, T. N. Nguyen, F. Massacci, and P. H. Phung, "HIT4Mal: Hybrid image transformation for malware classification," *Transactions on Emerging Telecommunications Technologies*, vol. 31, no. 11, Nov. 2020, doi: 10.1002/ett.3789.

[29] A. A. da Silva and M. Pamplona Segundo, "On Deceiving Malware Classification with Section Injection," *Mach. Learn. Knowl. Extr.*, vol. 5, no. 1, pp. 144–168, Jan. 2023, doi: 10.3390/make5010009.

[30] O. Aslan and R. Samet, "A Comprehensive Review on Malware Detection Approaches," 2020, *Institute of Electrical and Electronics Engineers Inc.* doi: 10.1109/ACCESS.2019.2963724.

[31] S. K. Smmarwar, R. Priyadarshi, P. Angaitkar, S. Mishra, and R. S. Rathore, "AIMD: AI-powered android malware detection for securing AIoT devices and networks using graph embedding and ensemble learning," *Journal of Systems Architecture*, vol. 173, p. 103707, Apr. 2026, doi: 10.1016/j.sysarc.2026.103707.

[32] M. Guven, "Leveraging deep learning and image conversion of executable files for effective malware detection: A static malware analysis approach," *AIMS Mathematics*, vol. 9, no. 6, pp. 15223–15245, 2024, doi: 10.3934/math.2024739.

[33] D. Xue, J. Li, T. Lv, W. Wu, and J. Wang, "Malware Classification Using Probability Scoring and Machine Learning," *IEEE Access*, vol. 7, pp. 91641–91656, 2019, doi: 10.1109/ACCESS.2019.2927552.

[34] P. Yadav, N. Menon, V. Ravi, S. Vishvanathan, and T. D. Pham, "EfficientNet convolutional neural networks-based Android malware detection," *Comput. Secur.*, vol. 115, p. 102622, Apr. 2022, doi: 10.1016/j.cose.2022.102622.

[35] Z. Zhang, P. Qi, and W. Wang, "Dynamic Malware Analysis with Feature Engineering and Feature Learning," *Proceedings of the AAAI Conference on Artificial Intelligence*, vol. 34, no. 01, pp. 1210–1217, Apr. 2020, doi: 10.1609/aaai.v34i01.5474.

[36] A. Vassilev, A. Oprea, A. Fordyce, and H. Anderson, "Adversarial machine learning: A Taxonomy and Terminology of Attacks and Mitigations," Jan. 2024. doi: 10.6028/NIST.AI.100-2e2023.

[37] J. Jeon, B. Jeong, S. Baek, and Y.-S. Jeong, "Static Multi Feature-Based Malware Detection Using Multi SPP-net in Smart IoT Environments," *IEEE Transactions on Information Forensics and Security*, vol. 19, pp. 2487–2500, 2024, doi: 10.1109/TIFS.2024.3350379.

[38] A. Damodaran, F. Di Troia, C. A. Visaggio, T. H. Austin, and M. Stamp, "A comparison of static, dynamic, and hybrid analysis for malware detection," *Journal of Computer Virology and Hacking Techniques*, vol. 13, no. 1, pp. 1–12, 2017, doi: 10.1007/s11416-015-0261-z.

[39] C. Tristan *et al.*, "Detecting Obfuscated Malware using Memory Feature Engineering (CIC-MalMem-2022)," in *The 8th International Conference on Information Systems Security and Privacy (ICISSP)*, 2022, pp. 177–188. doi: 10.5220/0010908200003120.

[40] S. M. Mathews, *Explainable Artificial Intelligence Applications in NLP, Biomedical, and Malware Classification: A Literature Review*, vol. 998. Springer International Publishing, 2019. doi: 10.1007/978-3-030-22868-2_90.

[41] O. Abdel Wahab, "Intrusion Detection in the IoT Under Data and Concept Drifts: Online Deep Learning Approach," *IEEE Internet Things J.*, vol. 9, no. 20, pp. 19706–19716, Oct. 2022, doi: 10.1109/JIOT.2022.3167005.

[42] J. Tripathi, H. Gomes, and M. Botacin, "Towards Explainable Drift Detection and Early Retrain in ML-Based Malware Detection Pipelines," 2025, pp. 3–24. doi: 10.1007/978-3-031-97623-0_1.

[43] X. Zhang *et al.*, "Detecting and Adapting to Stealthy Label-Inversion Drifts via Conditional Distribution Inference," in *Proceedings - 28th International Symposium on Research in Attacks, Intrusions and Defenses, RAID 2025*, Institute of Electrical and Electronics Engineers Inc., 2025, pp. 204–219. doi: 10.1109/RAID67961.2025.00052.

[44] Y. Dehfouli and A. H. Lashkari, "VADViT: Vision transformer-driven memory forensics for malicious process detection and explainable threat attribution," *Journal of Information Security and Applications*, vol. 94, p. 104200, Nov. 2025, doi: 10.1016/j.jisa.2025.104200.

[45] L. Yang, A. Ciptadi, I. Laziuk, A. Ahmadzadeh, and G. Wang, "BODMAS: An Open Dataset for Learning based Temporal Analysis of PE Malware," in *2021 IEEE Security and Privacy Workshops (SPW)*, IEEE, May 2021, pp. 78–84. doi: 10.1109/SPW53761.2021.00020.

[46] A. Habibi Lashkari, M. Shafi, Y. Li, A. P. Singh, and A. Barkworth, "Unveiling evasive malware behavior: toward generating a multi-sources benchmark dataset and evasive malware behavior profiling using network traffic and memory analysis," *J. Supercomput.*, vol. 81, no. 6, p. 782, Apr. 2025, doi: 10.1007/s11227-025-07267-x.

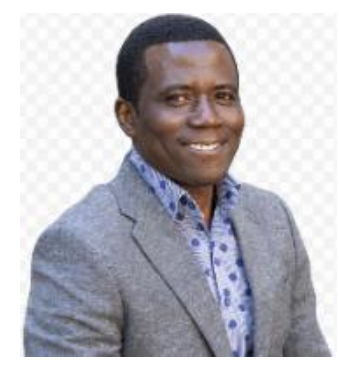

**Isaac K. Nti** (Member, IEEE) received the Ph.D. degree in Computer Science from the University of Energy and Natural Resources, Ghana, in 2021, and the M.Sc. degree in Information Technology from Kwame Nkrumah University of Science and Technology, Ghana, in 2016. He is with the School of Information Technology, University of Cincinnati, USA, and is affiliated with the Information Technology and Analytics Center. His research interests include applied machine learning for cybersecurity, explainable and trustworthy AI, adversarial machine learning, and lightweight models for edge and cyber-physical systems. He is also a Full Member of Sigma Xi.